\documentclass[usenatbib]{mn2e}
\usepackage{graphicx}
\usepackage{epstopdf}
\usepackage{natbib}
\usepackage{verbatim}
\usepackage{amssymb,amsmath}
\usepackage{ulem}
\usepackage{color}
\usepackage{comment}

\renewcommand{\thefootnote}{\fnsymbol{footnote}} 
\newcommand\aj{AJ}
\newcommand\araa{ARA\&A}
\newcommand\apj{ApJ}
\newcommand\apjl{ApJ}
\newcommand\apjs{ApJS}
\newcommand\aap{A\&A}
\newcommand\mnras{MNRAS}
\newcommand\pasa{PASA}
\newcommand\pasj{PASJ}
\newcommand\nat{Nature}

\newcommand{\mg}{MgII}
\newcommand{\fe}{FeII}

\newcommand{\Msun}{${\rm M_\odot}$}

\title[Fe/Mg vs Redshift]{On the Cosmic Evolution of Fe/Mg in QSO Absorption Line Systems}

\author[Dey et al.]
       {\parbox{18cm}{
       Arjun Dey$^{1,2}$\footnotemark[1], 
       Paul Torrey$^{3,4}$,
       Kate H. R. Rubin$^{5}$,	
       Guangtun Ben Zhu$^{6,7}$,
       Joshua Suresh$^{5}$
       }\vspace{0.3cm}\\ 
         $^1$ National Optical Astronomy Observatory, Tucson, AZ 85726\\ 
         $^2 $ Fellow, Radcliffe Institute for Advanced Study, Harvard University, Byerly Hall, 8 Garden Street, Cambridge, MA 02138\\
	 $^3 $ TAPIR 350-17, California Institute of Technology, 1200 E. California Boulevard, Pasadena, CA 91125, USA\\
         $^4$ MIT Kavli Institute for Astrophysics \& Space Research, Cambridge, MA, 02139\\
         $^5 $ Harvard-Smithsonian Center for Astrophysics, 60 Garden Street, Cambridge, MA 02138\\
         $^6$ Department of Physics \& Astronomy, Johns Hopkins University, 3400 N. Charles Street, Baltimore, MD 21218, USA\\
 	$^7$ Hubble Fellow\\
         }

\begin{document}
\maketitle

\begin{abstract}
We investigate the variation of the ratio of the equivalent widths of the 
FeII$\lambda$2600 line  to the 
MgII$\lambda\lambda$2796,2803 doublet 
as a function of redshift in a large sample of absorption lines drawn from the Johns Hopkins University - Sloan Digital Sky Survey Absorption Line Catalog.
We find that despite large scatter, the observed ratio shows a trend where the equivalent width ratio $\mathcal{R}\equiv W_{\rm FeII}/W_{\rm MgII}$ decreases monotonically with increasing redshift $z$ over the range $0.55 \le z \le 1.90$. Selecting the subset of absorbers where the signal-to-noise ratio of the MgII equivalent width $W_{\rm MgII}$ is $\ge$3 and modeling the equivalent width ratio distribution as a gaussian, we find that the mean of the gaussian distribution varies as $\mathcal{R}\propto (-0.045\pm0.005)z$. We discuss various possible reasons for the trend. A monotonic trend in the Fe/Mg abundance ratio is predicted by a simple model where the abundances of Mg and Fe in the absorbing clouds are assumed to be the result of supernova ejecta and where the cosmic evolution in the SNIa and core-collapse supernova rates is related to the cosmic star-formation rate. If the trend in $\mathcal{R}$ reflects the evolution in the abundances, then it is consistent with the predictions of the simple model. 
\end{abstract}


\renewcommand{\thefootnote}{\fnsymbol{footnote}}

\footnotetext[1]{E-mail: dey@noao.edu}


\section{Introduction}


The absorption lines of heavy elements (e.g., C, Mg, O, Fe) observed in the spectra of distant quasi-stellar objects (QSOs) provide one of the best ways of observing and understanding the evolving chemical enrichment of the Universe. Several lines of evidence suggest that these absorption lines arise in the circumgalactic medium (CGM) surrounding 
galaxies lying close to the lines of sight to QSOs 
\citep[e.g.,][]{Bahcall1969,Yanny1990,Bergeron1991,Steidel1995,Zhu2014}. 
The source of the heavy elements in the CGM is not clear, but 
it is thought to result from enrichment of the halo gas by large-scale galactic outflows driven by a combination of stellar winds, supernovae, active galactic nuclei (AGN) and by the accretion of gas tidally stripped during mergers from star-forming galactic disks \citep[e.g.,][]{Oppenheimer2006,DallaVecchia2008,Suresh2015}.
The ubiquity of galactic winds associated with star-formation in galaxies over a wide range in redshift is 
now empirically well established \citep[e.g.,][]{Weiner2009,Erb2012,Martin2012,Kornei2012,Rubin2014} by the detection of blue-shifted absorption lines of MgII and FeII in galaxies. These outflows can carry enriched material out of galactic disks to distances of $\ga$50~kpc into galaxy halos \citep{Bordoloi2011,Rubin2014} and thus enrich the CGM \citep[e.g.,][]{Prochaska2011}. Outflows of enriched material from galaxy disks have also been invoked to regulate the mass-metallicity relationship observed in galaxies \citep[e.g.,][]{Tremonti2004,Finlator2008,Dave2011,Torrey2014}: star-formation, fuelled by accreted gas, can increase a galaxy's metallicity, but outflows can remove enriched gas from the galaxy and, in lower mass systems, prevent further star-formation and enrichment \citep[e.g.,][]{Mannucci2010}. 

In this picture the metal enrichment of the circumgalactic medium is the result of supernovae  and a natural consequence is that the ratio of $\alpha$ elements to Fe should reflect (to first order) the relative rates of core collapse supernovae (CCSN; which result from the explosion of a massive star) and Type Ia supernovae (SNIa; which result from the explosion of an accreting white dwarf). The heavy element yields from both classes of supernovae have been modeled in various studies \citep{Nomoto1997,Thielemann2003}, which typically find that the $\alpha$-elements are formed predominantly in CCSN, whereas Fe production is dominated by SNIa. The observed ratio of $\alpha$-element to Fe equivalent widths in the spectra of absorption line systems over a range of redshifts can potentially provide an important window into the relative variation of SN rates, the efficiency of outflows in enriching the CGM, 
and the resultant abundance patterns of diffuse CGM and IGM material
with cosmic time. 
 
In this paper, we use the Johns Hopkins University - Sloan Digital Sky Survey (JHU-SDSS) Metal Absorption Line Catalog \citep[][hereafter ZM13]{ZM13} to investigate the variation of the FeII/MgII ratio as a function of redshift. The catalog tabulates the equivalent widths for  35,752 absorption line systems measured in the spectra of QSOs observed by the Sloan Digital Sky Survey DR7 data release \citep{sdssdr7,York2000}. We focus on the measurements of the equivalent widths of the MgII$\lambda\lambda$2796.35,2803.53 ($^2S\rightarrow  {^2P}_{{{1}\over{2}},{{3}\over{2}}}$) and FeII$\lambda\lambda$2586.00,2600.73 ($a\, {^6D}_{{9}\over{2}}\rightarrow z\, {^6D}_{{{7}\over{2}},{{9}\over{2}}}$) transitions \citep[see, e.g.,][]{Prochaska2011,Hartigan1999}. 

We present the sample in \S2 and our analyses in \S3 and show that there is evidence for a trend of decreasing $W_{\rm FeII}/W_{\rm MgII}$ with redshift. In \S4 we discuss the possible origins of this trend. We consider the caveats and whether the ratio is a viable proxy for understanding the abundance ratio of Fe to Mg. We suggest that the observed trend reflects the possible cosmic evolution of the Fe/Mg ratio and is consistent with the predictions of a simple model where the abundance ratio variations are driven by the ratio of the rates of SNIa to CCSN. 

Throughout this paper, we adopt a cosmology with [$\Omega_m,\Omega_\Lambda,\sigma_8,h$] = [0.3,0.7,0.8,0.7]. 
Magnitudes are presented in AB units \citep{ABmag} unless otherwise specified. 

\begin{figure*}
\centerline{\vbox{\hbox{
\includegraphics[width=7.3in]{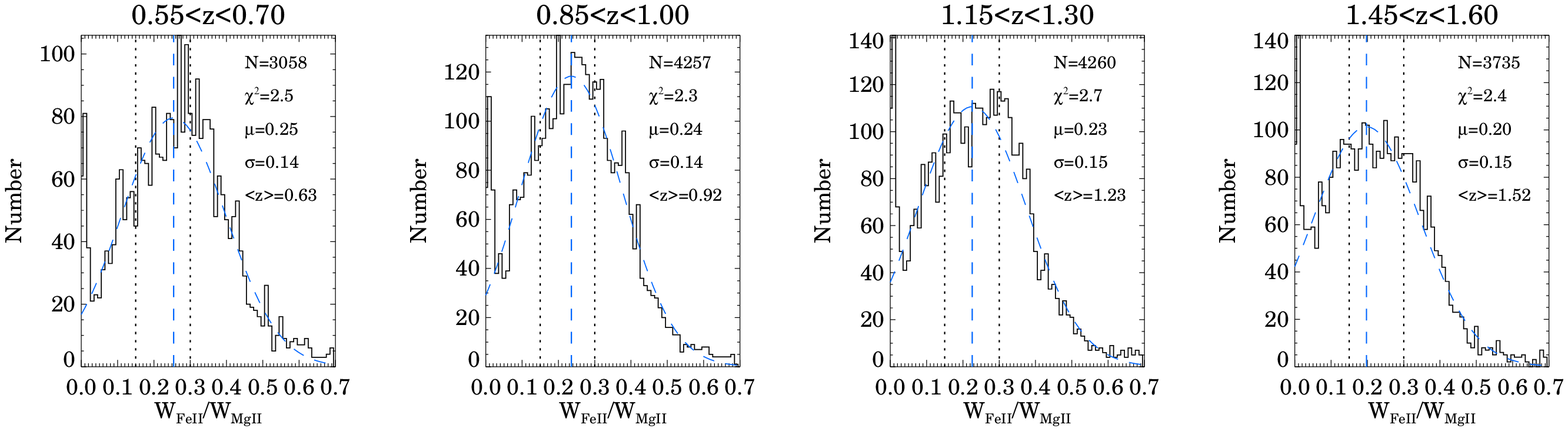}
}}}
\caption{The histograms show sample distributions of the equivalent width ratios in four different redshift bins. The vertical dotted lines at $W_{\rm FeII}/W_{\rm MgII}$ = 0.15 and 0.30 are shown as fiducials. The gaussian fits discussed in the text are shown as dashed curves with the peak value shown by the vertical dashed line. The legend in each panel lists the number of absorbers in the redshift bin (N), the $\chi^2$ value of the gaussian fit, the mean ($\mu$) and standard-deviation ($\sigma$) of the best-fit gaussian, and the median redshift ($<z>$) of the absorbers within the bin. 
\label{histograms}}
\end{figure*}

\section{Sample Definition}

In this work, we make use of the JHU-SDSS Metal Absorption Line Catalog \citep[][; hereafter ZM13]{ZM13} based on the SDSS DR7 release \citep{sdssdr7}\footnote{http://www.pha.jhu.edu/~gz323/jhusdss}.

This catalog of 35,752 MgII absorption line systems is constructed from 84,533 quasar spectra from SDSS DR7 and covers the redshift range between about 0.4 and 2.2 (determined by the wavelength coverage of the SDSS spectrographs; York et al. 2000). The window function, completeness and purity of the catalog were determined and cross-checked using different approaches, including Monte Carlo simulation for the whole sample and redshift path determination for individual quasars (see ZM13 for details). 
The statistical incidence rate of the Mg~II doublet as a function of redshift and strength was also presented in ZM13 and compared with the literature.

The systematic effects in the catalog are also well understood. The primary systematic effects are: (1) the imperfect calibration of the relative spectrophotometry, due to imperfect F-star SED models; (2) the absorption induced by ISM/CGM of the Milky Way at given observer-frame wavelengths, most importantly at 3934\AA\ and 3969\AA\ (due to Ca~II), 5891\AA\ and 5897\AA\ (due to Na I), and various diffuse interstellar bands; (3) poor subtraction of the telluric emission lines, which were, on average, under-subtracted in the reduction pipeline; and (4) relatively low completeness at both the short and long wavelength ends due to the reduced throughput at the blue end of the SDSS spectrographs and the poor subtraction of the bright OH lines at the red end.  Finally, comparisons of the ZM13 catalog to an independently constructed DR7 catalog find good agreement in the measured equivalent widths in the two catalogs, with the difference in $W_{\rm MgII}$ measurements less than $0.02$~\AA\ \citep{Seyffert2013, Gauthier2014}.

The ZM13 catalog therefore represents a well-understood dataset that spans a large range in redshift; the known systematics have negligible effects on any of the statistics derived here. 
The spectrophotometric calibration errors are at $\approx 10^{-3}$ level (about an order of magnitude smaller than the photon Poisson noise) and therefore have no effect on the statistics presented here. The absorption by Galactic Ca II can be strong along some lines of sight and does affect the detection efficiency of extragalactic Mg II absorbers below redshift 0.43. We therefore choose not to include the absorbers at $z<0.43$ in our sample. In addition, we restrict the fits and statisitical analyses to $z>0.55$ to ensure all detected FeII$\lambda$2600 lines lie in the region of higher completeness at $\lambda>4000$\AA. The terrestrial lines are known to be under-subtracted and the noise (i.e., flux errors) at corresponding wavelengths were adjusted in accordance in the SDSS reduction pipeline. The Mg II detection pipeline has been fully-tested on cases with poor sky subtraction and shown to perform at the same efficiency as along other sightlines. These tests are discussed in detail in \citet{ZM13} and \citet{Lan2014}. At the red end, we restrict our analyses to $z<1.90$, which corresponds to MgII detections at roughly $\lambda<8100$\AA.

We 
extracted all measurements of the \mg\ $\lambda\lambda$2796.35,2803.53, MgI$\lambda$2853,  \fe\ $\lambda\lambda$2586.00,2600.73 absorption lines. The catalog contains a total of 35,752 absorption line systems, $>$97\% of which are ``strong" absorbers with $W_{\rm MgII2796}>0.3$\AA. Since a few of these are only detected marginally, we restricted our analyses to a subset of 35,049 systems with absorption line signal-to-noise ratios of ${\rm SNR_{\rm MgII}\equiv W_{\rm MgII}/\sigma_{W_{\rm MgII}}\ge3}$, where $W_{\rm MgII}$ is the sum of the equivalent widths of the \mg\ 2796.35 and 2803.53 resonance lines and $\sigma_{W_{\rm MgII}}$ is the measurement uncertainty; 99.5\% of these have $W_{\rm MgII2796}>0.3$. No restriction was placed on the strength of the FeII line (hereafter \fe\ refers to the 2600.73 line). 31,727 of the systems lie at $0.55\le z \le 1.90$ and constitute our primary statistical sample.

Some data in the catalog required editing. 2,554 sources have values of $W_{\rm FeII}\le 10^{-3}$\AA, 816 of which lie at $z<0.55$ (and therefore outside our main sample) and the rest of which are distributed in redshift similarly with the overall sample of absorbers. 7 more sources have $\sigma_{W_{\rm FeII}}\ge 5$\AA\ (most with meaningless values of many 1000s) and 3 have $\sigma_{W_{\rm FeII}}=0$. In all these systems, we considered FeII to be undetected and, for the purposes of our analyses, we set $W_{\rm FeII}=0.015$\AA, the minimum detectable equivalent width in the catalog, and the uncertainty $\sigma_{W_{\rm FeII}}=\sigma_{W_{\rm MgII}}$. In all cases, this resulted in $\sigma_{W_{\rm FeII}}/W_{\rm FeII}>1.65$. Within $0.55\le z \le 1.90$, the redshift distribution of the $<1\sigma$ $W_{\rm FeII}$ is very similar to that of the $>1\sigma\,\,W_{\rm FeII}$ equivalent widths.

\section{Results and Analysis}

For each absorber, we defined the ratio $\mathcal{R}\equiv W_{\rm FeII}/W_{\rm MgII}$. 
The distribution of $\mathcal{R}$ shows a broad distribtion which is roughly gaussian with a weak tail to high values of $\mathcal{R}$. Of the 31,727 absorbers, 4,965 (15.6\%) have $\mathcal{R}\le \sigma_\mathcal{R}$. The redshift distribution of these sources is similar to the overall distribution of absorbers.

Given the large scatter in the points and the non-detections of FeII, we considered multiple approaches to determining whether there is a trend of $\mathcal{R}$ with redshift and estimating its significance. 

\subsection{Fits to the $W_{\rm FeII}/W_{\rm MgII}$ Point Distribution}

A least-absolute deviation fit to the overall point distribution using 1000 bootstrap samples (with replacement) results in a median slope of $\Delta\mathcal{R}/\Delta z = -0.046\pm0.003$. The 95\% confidence limits on the slope are [-0.040,-0.052]. 

Given the significant fraction of FeII non-detections, we also measured the linear regression using the expectation-maximization (EM) technique and Schmitt's binned linear regression method, as implemented in the ASURV package \citep{ASURV1992,ASURVBIVAR}. All points with $\mathcal{R}\le \sigma_\mathcal{R}$ were treated as upper limits. The EM method resulted in a slope estimate of $-0.0406\pm0.0035$. Schmitt's method was used with a redshift binning of $\delta(z)=0.15$ and $\mathcal{R}$ binning of $\delta(\mathcal{R})=0.05$ and resulted in a slope estimate of $-0.034$. We investigated different bin sizes, but the results did not change significantly. 

Finally, we used the Bayesian approach to linear regression developed by \citet{Kelly2007}. This was very CPU-intensive and slow for our sample size, but resulted in a median slope of $-0.0304$ with a 95\% confidence range of [$-0.0343,-0.0252$]. 


\subsection{Fits to the Shape of the $W_{\rm FeII}/W_{\rm MgII}$ Distribution}

We next investigate the redshift evolution in the shape of the distribution of $\mathcal{R}\equiv W_{\rm FeII}/W_{\rm MgII}$. In this approach, non-detections were just placed at their ``measured'' or edited value; given that the non-detections follow the redshift distribution of the detected sample, any widening of the overall distribution is likely to be similar in all redshift bins. 
 
Following ZM13, we divided the sample into 12 redshift bins defined between the redshifts 
$z$=[0.43,0.55,0.70,0.85,1.00,1.15,...,2.05,2.30] and 
measured the distribution of the ratio $\mathcal{R}\equiv W_{\rm FeII}/W_{\rm MgII}$ in each bin. We modeled the distribution in each redshift bin as a gaussian (fit using a non-linear least squares method) and found the best fit value for the gaussian mean. We also computed the median value of the distribution within each bin. Fig.~\ref{histograms} shows the distributions and the fits in a few sample redshift bins. We then performed linear fits (using least absolute deviation and least squares techniques) to the median and mean values to measure the variation of $\mathcal{R}$ with redshift.

In order to understand the uncertainty that may be inherent in this analysis due to the effects of a complex and censored sample, we carried out a bootstrap analysis by selecting samples using replacement from the data in each redshift bin. For each bootstrap sample, we carried out the same analyses (i.e., measuring medians, fitting gaussians, and fitting the median and means as a linear function of redshift). 

\begin{figure}
\centerline{\vbox{\hbox{
\includegraphics[width=3.5in]{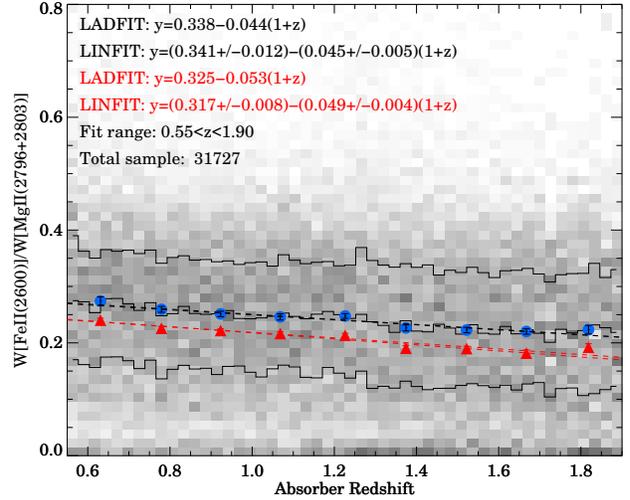}
}}}
\caption{Distribution of the ratio of the \fe\ $\lambda$2600 to \mg\ $\lambda\lambda$2796,2803 equivalent widths as a function of redshift. The greyscale show the distribution of all absorption line systems from the JHU-SDSS Absorption Line Catalog \citep{ZM13} with $>3\sigma$ $W_{\rm MgII}$ measurements. The thin black lines show the median and quartiles of the distribution. 
The solid blue circles show the median values in different redshift bins; the solid pink triangles show the peak of gaussian fits to the ratio distribution (see text). The dashed lines show linear fits to the two sets of solid points. The grey hashed region denotes the range of slopes resulting from the Monte Carlo simulations of the null hypothesis (i.e., intrinsic $\mathcal{R}$ constant with redshift) described in the text. 
A small fraction (1.4\%) of the absorbers have ratios larger than the y-limit shown. We include them in our analyses, but they have no effect on the results. 
\label{absratioplot}}
\end{figure}

The results for the primary sample (i.e., 31,727 $W_{\rm MgII}/\sigma_{W_{\rm MgII}}\ge 3$ absorbers) are shown in Figure~\ref{absratioplot}. The distribution of points is shown as a greyscale with the median and quartiles represented as thin black lines. 
The best fit peak and median values in each redshift bin are shown as large filled triangles and circles, respectively, along with the uncertainties derived in the bootstrap analyses. The peak and median values of the $\mathcal{R}$ ratio distribution trend to lower values with increasing redshift. The same behavior is reflected in the variation of the quartiles with redshift. A linear fit to the peak values of the gaussian distribution within the redshift range $0.55\le z\le1.9$ (i.e., using redshift bins with $>$900 absorbers) results in $\mathcal{R}\propto(-0.049\pm 0.006)z$ with 95\% confidence limits of [$-0.055,-0.036$]; a fit to the median values results in a similar slope of $(-0.045\pm0.005)$. 


The observed trend is not due to the censorship of the data. For our primary sample we used the least censored subset. However, we also investigated subsamples selected with a range of signal-to-noise ratios ${\rm SNR_{\rm FeII or MgII}\ge [0,1,2,3,5,7,10]}$ and found significant trends (with $\mathcal{R}$ decreasing with increasing redshift in all cases) and with slopes (derived for fits to the median values) ranging from $-0.030$ to $-0.056$, with a median derived slope value of $\approx -0.045$. The steepest (shallowest) slope values are found for the ${\rm SNR_{\rm MgII}\ge7}$ and ${\rm SNR_{\rm FeII}\ge0}$ (${\rm SNR_{\rm MgII}\ge10}$ and ${\rm SNR_{\rm FeII}\ge10}$) sample containing 20,813 (2,962) absorbers (see Table~1).

\begin{table*}
\begin{tabular}{cccccccc}
\hline
 & \multicolumn{7}{c}{SNR(FeII)} \\
 {SNR(MgII)} & $\ge$0  &$\ge$1  & $\ge$2  & {$\ge$3}  & {$\ge$5}  & {$\ge$7} & {$\ge$10} \\
\hline
   $\ge$0 &  -0.0430$\pm$0.0048 &  -0.0506$\pm$0.0046 &  -0.0590$\pm$0.0044 &  -0.0522$\pm$0.0060 &  -0.0417$\pm$0.0046 &  -0.0355$\pm$0.0061 &  -0.0325$\pm$0.0069\\
 &  -0.0468$\pm$0.0044 &  -0.0678$\pm$0.0034 &  -0.0569$\pm$0.0030 &  -0.0522$\pm$0.0027 &  -0.0432$\pm$0.0030 &  -0.0372$\pm$0.0033 &  -0.0339$\pm$0.0043\\
            &  32167 &  27042 &  20487 &  14873 &   8687 &   5483 &   3044 \\
 & & & & & & & \\
   $\ge$1 &  -0.0428$\pm$0.0050 &  -0.0505$\pm$0.0047 &  -0.0586$\pm$0.0044 &  -0.0524$\pm$0.0056 &  -0.0417$\pm$0.0046 &  -0.0353$\pm$0.0062 &  -0.0331$\pm$0.0071\\
 &  -0.0470$\pm$0.0043 &  -0.0674$\pm$0.0034 &  -0.0570$\pm$0.0030 &  -0.0525$\pm$0.0027 &  -0.0431$\pm$0.0030 &  -0.0373$\pm$0.0034 &  -0.0339$\pm$0.0042\\
            &  32157 &  27034 &  20483 &  14869 &   8686 &   5483 &   3044 \\
 & & & & & & & \\
   $\ge$2 &  -0.0432$\pm$0.0048 &  -0.0508$\pm$0.0047 &  -0.0590$\pm$0.0045 &  -0.0527$\pm$0.0056 &  -0.0420$\pm$0.0047 &  -0.0351$\pm$0.0063 &  -0.0325$\pm$0.0066\\
 &  -0.0468$\pm$0.0042 &  -0.0673$\pm$0.0035 &  -0.0569$\pm$0.0028 &  -0.0525$\pm$0.0028 &  -0.0434$\pm$0.0031 &  -0.0372$\pm$0.0034 &  -0.0338$\pm$0.0041\\
            &  32137 &  27021 &  20474 &  14866 &   8686 &   5483 &   3044 \\
 & & & & & & & \\
   $\ge$3 & {\bf -0.0445$\pm$0.0049} &  -0.0496$\pm$0.0046 &  -0.0583$\pm$0.0042 &  -0.0516$\pm$0.0053 &  -0.0417$\pm$0.0047 &  -0.0351$\pm$0.0061 &  -0.0319$\pm$0.0068\\
 &  {\bf -0.0487$\pm$0.0045} &  -0.0667$\pm$0.0034 &  -0.0568$\pm$0.0030 &  -0.0524$\pm$0.0028 &  -0.0432$\pm$0.0030 &  -0.0370$\pm$0.0034 &  -0.0336$\pm$0.0040\\
            &  {\bf 31727} &  26864 &  20447 &  14851 &   8679 &   5478 &   3043 \\
 & & & & & & & \\
   $\ge$5 &  -0.0543$\pm$0.0051 &  -0.0478$\pm$0.0045 &  -0.0521$\pm$0.0047 &  -0.0503$\pm$0.0056 &  -0.0410$\pm$0.0046 &  -0.0333$\pm$0.0061 &  -0.0315$\pm$0.0069\\
 &  -0.0582$\pm$0.0042 &  -0.0634$\pm$0.0036 &  -0.0525$\pm$0.0028 &  -0.0520$\pm$0.0028 &  -0.0430$\pm$0.0030 &  -0.0370$\pm$0.0034 &  -0.0336$\pm$0.0043\\
            &  27602 &  24397 &  19655 &  14663 &   8647 &   5457 &   3035 \\
 & & & & & & & \\
   $\ge$7 &  -0.0578$\pm$0.0043 &  -0.0487$\pm$0.0054 &  -0.0431$\pm$0.0045 &  -0.0481$\pm$0.0048 &  -0.0416$\pm$0.0045 &  -0.0337$\pm$0.0059 &  -0.0308$\pm$0.0069\\
 &  -0.0608$\pm$0.0037 &  -0.0599$\pm$0.0034 &  -0.0473$\pm$0.0030 &  -0.0474$\pm$0.0026 &  -0.0438$\pm$0.0029 &  -0.0372$\pm$0.0035 &  -0.0340$\pm$0.0042\\
            &  20813 &  19357 &  16896 &  13619 &   8495 &   5396 &   3008 \\
 & & & & & & & \\
  $\ge$10 &  -0.0563$\pm$0.0078 &  -0.0499$\pm$0.0064 &  -0.0459$\pm$0.0048 &  -0.0441$\pm$0.0042 &  -0.0427$\pm$0.0051 &  -0.0364$\pm$0.0054 &  -0.0305$\pm$0.0067\\
 &  -0.0610$\pm$0.0039 &  -0.0601$\pm$0.0039 &  -0.0491$\pm$0.0033 &  -0.0455$\pm$0.0029 &  -0.0448$\pm$0.0030 &  -0.0410$\pm$0.0032 &  -0.0353$\pm$0.0043\\
            &  13483 &  13028 &  12234 &  10879 &   7859 &   5222 &   2962 \\
\hline
\label{censortab}
\end{tabular}
\caption{Trend slopes for subsamples defined by censoring the data by the signal-to-noise ratio in $W_{\rm FeII}$ and $W_{\rm MgII}$. For each pair of signal-to-noise ratio cuts, the top entry is the median slope and 1$\sigma$ error (determined using bootstrap resampling) determined by fitting (using least absolute deviation) the median values in each redshift bin; the middle entry is the median and 1$\sigma$ error for the slope determined by a linear least squares fit to the mean of the gaussian model for the distribution; and the bottom entry is the total number of absorbers in the censored sample. The primary sample used in this paper is the one listed in bold face for SNR(Mg)$\ge$3 and SNR(Fe)$\ge$0. }
\end{table*}

While higher SNR thresholds tend to select higher equivalent width lines, the correspondence between the two quantities is complicated and the equivalent width distribution at a given signal-to-noise threshold varies in a complicated way with wavelength. Hence, in order to further investigate whether the trend is dominated (or damped) by stronger absorbers, we investigated subsamples with a range of $W_{\rm MgII}$ and $W_{\rm FeII}$ equivalent width cutoffs, selecting samples with equivalent width cuts $>$ [0,0.5,1.0,1.5,2.0] in either line. The derived slopes do not vary in a monotonic way with the equivalent width cuts, and there is no evidence from the current dataset that a subset of stronger or weaker absorbers are dominating the trend signal (see also Fig.~\ref{saturation} right panel).

The trend of $\mathcal{R}$ with redshift appears to be primarily driven by an evolution in the \fe\ equivalent width. Fitting the medians of the distributions of the \mg\ and \fe\ equivalent widths as a function of redshift results in measured slopes of $-$0.06$\pm$0.05 for $W_{\rm MgII}$ vs. $z$ and $-$0.20$\pm$0.02 for $W_{\rm FeII}$ vs. $z$.  The non-evolution in the slope of $W_{\rm MgII}$ observed in our data is consistent with the results from previous high-spectral-resolution studies \citep{Mshar2007,Churchill2003}. 

\subsection{Are Selection Effects Responsible for the Trend?}

We have established that the absorption line systems listed in the JHU-SDSS catalog exhibit a trend where the $W_{\rm FeII}/W_{\rm MgII}$ decreases with redshift. Although the catalog is constructed from the spectra of QSOs observed uniformly with the SDSS spectrographs, it is important to consider whether the observed trend could be the result of observational selection effects. 

As an initial test, we investigated whether the line ratios \mg\ $\lambda$2796/\mg\ $\lambda$2803 and \fe~$\lambda$2586/\fe~$\lambda$2600 in our sample show significant variation with redshift; they do not, and fits (to SNR$\ge$3 samples) result in slopes of $-$0.01$\pm$0.13 and $-$0.03$\pm$0.01 respectively (see also the left two panels of Fig.~\ref{saturation}). In addition, the MgI$\lambda$2853/MgII equivalent width ratio does not vary significantly with redshift (slope of 0.02$\pm$0.01). Changing the subsamples by varying the signal-to-noise ratio cutoff does not change these results significantly, although in the case of the \fe~$\lambda$2586/\fe~$\lambda$2600 ratio, higher SNR cuts result in a slightly more significant detection of a trend (e.g., for SNR$>$5, \fe~$\lambda$2586/\fe~$\lambda$2600 vs. redshift has a slope of $-0.045\pm0.010$).
While the wavelength separation between these various features is smaller than that between the \mg\ and \fe\ features, there does not appear to be any systematic effect introduced by the fact that the lines are measured in different parts of the spectrum. The division in spectral coverage between the blue and red spectrographs at $\lambda\sim5800$\AA\ does result in a lower detection efficiency of absorption lines at these wavelengths and results in a slightly higher median equivalent width for the $z\approx 1.2$ bin; however, there is no obvious feature in the distribution of the line ratios and this bin does not affect the determination of the overall trend.  

In order to determine whether the complicated selection function for absorption line detection could be responsible for the trend, we used the measured selection function from ZM13 to create 1000 simulated datasets as follows. We simulated a large number of \mg\ absorption lines (distributed randomly in redshift within each redshift bin) using the best-fit intrinsic exponential distribution of rest equivalent width computed by ZM13 (see equation 3 and table 1 in ZM13). For each \mg\ absorber, the $W_{\rm MgII2796}/W_{\rm MgII2803}$  doublet ratio was drawn from a lognormal model based on a fit to the observed distribution of the ratio in the JHU-SDSS dataset. Then, for each \mg\ absorber, an \fe\ absorption line was drawn randomly from a normal distribution of $\mathcal{R}$ with a mean of 0.233 and $\sigma$=0.143. In order to test whether the selection would generate a systematic bias, the $\mathcal{R}$ distribution was assumed to be constant with redshift. 
Equivalent width uncertainties for the MgII and FeII lines were assigned randomly, drawn from model fits to the actual fractional error distributions in these two lines.  
In each case, we constructed the two-dimensional distribution of $\sigma_W/W$ versus $W$ and modeled the surface as a lognormal in $\sigma_W/W$ and a polynomial in $W$. 
We then ``observed" this set of intrinsic \mg+\fe\ absorbers using the equivalent width selection function for the JHU-SDSS catalog (ZM13). We selected absorbers and fit each of the 1000 such datasets in the same manner in which we fit the observed catalog and derived slope measurements for any gradient in $W_{\rm FeII}/W_{\rm MgII}$ ratio with redshift. Since there was no evolution assumed in the data, any resulting slope is purely due to selection effects (i.e., completeness as a function of equivalent width and observed wavelength) and the variation in the intrinsic distribution of $W_{\rm MgII}$ with redshift. 

We do not detect the observed slope in the simulated $\mathcal{R}$ distribution with a high degree of reliability. The 1000 trials resulted in a normal distribution of slopes with a small negative bias (median recovered slope = $-0.0148$; standard deviation of 0.0033); the 95\% (99\%) confidence limits on the slope are [$-0.021,-0.008$] ([$-0.023,-0.0062$]) and none of the 1000 trials resulted in slope determinations $\le-0.0245$. 
We therefore conclude that the observed trend is not the result of selection effects. 

\section{Discussion}

Our results show that the ratio of the equivalent widths of \fe\ to \mg\ absorption lines along the lines of sight to QSOs decreases with redshift. The trend survives a range of signal-to-noise ratio or equivalent width cuts and does not appear to be due to selection effects within the JHU-SDSS ZM13 dataset. In this section, we discuss possible physical origins for the equivalent width ratio evolution. 

The strength of an isolated absorption line as measured by its equivalent width is related to its abundance in the gas phase and the velocity spread of the gas \citep{Menzel1936,Baker1936,Aller1942}.  It is tempting to interpret the $W_{\rm FeII}/W_{\rm MgII}$ ratio in terms of the Fe/Mg abundance ratio. However, in order to interpret the observed equivalent widths in terms of physical quantities associated with the absorbing clouds, we need to consider the effects of (i) saturation, (ii) kinematics, (iii) ionization, and (iv) depletion. We consider each of these in turn.



\noindent{\it (i) Line Saturation:}
The $W_{\rm MgII}$ distribution for the SNR$\ge$3 sample is a lognormal which peaks at $\approx2$\AA, implying that the individual doublet lines have typical equivalent widths $>$1\AA\ and lie in the logarithmic (i.e., ``saturated") part of the curve of growth. It is therefore not obvious that we can interpret any observed trend with redshift as resulting from a trend in the abundance ratio of Mg to Fe without first understanding the effects of line saturation. 

In order to investigate the effects of saturation on the observed ratio, we computed the curves of growth for \mg\ and \fe\ for two different values of the Doppler parameter, $b_D$=10~km~s$^{-1}$ and 50~km~s$^{-1}$. For each case, we computed the expected $\mathcal{R}$ values that result from a wide range in \mg\ column density $12\le {\rm log}(N_{\rm MgII}) \le 17.9$ and [Mg/Fe] varying between 1 and 10 times the solar value. For a given $b_D$, even strongly saturated lines result in the equivalent width ratio $\mathcal{R}$ depending on the [Mg/Fe] abundance ratio. Hence, even if the observed absorption lines are single and saturated, they only dampen, but not erase, an intrinsic variation of the abundance ratio. Most MgII absorption systems with $W_{\rm MgII}>0.3$\AA\ are multiple systems \citep[e.g.,][]{Churchill2003}, and this complicates the simple picture described here. Fully modeling the result of saturation and multiplicity is beyond the scope of this paper, but instead we look to the data for clues to whether saturation effects are responsible for the observed trends. 

\begin{figure*}
\centerline{\vbox{\hbox{
\includegraphics[width=7.3in]{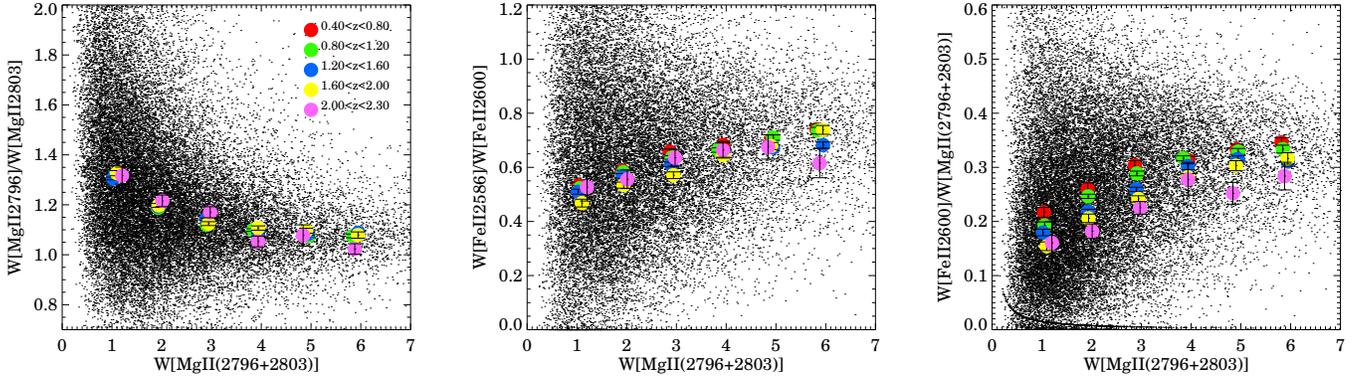}
}}}
\caption{The impact of line saturation, shown as the variation of $W_{\rm MgII2796}/W_{\rm 2803}$ (Left panel), $W_{\rm FeII2586}/W_{\rm Fe2600}$ (Middle panel), and $W_{\rm FeII}/W_{\rm MgII}$ (Right panel)  with the \mg\ equivalent width. The small points show the distributions for the $\ge3\sigma$ sample described in the text and the large coloured circles show the median values of the distributions in the redshift bins shown in the legend. The left two panels show that the ratio asymptotes to a constant value with increasing $W_{\rm MgII}$, suggesting that the lines are becoming saturated, but that the behavior is the same at all redshifts. In contrast, the $W_{\rm FeII}/W_{\rm MgII}$ (Right panel) shows 
a systematic evolution with redshift in all $W_{\rm MgII}$ bins. 
\label{saturation}}
\end{figure*}

Fig.~\ref{saturation} shows the effect of saturation on three pairs of equivalent width ratios in our observational dataset. The $W_{\rm MgII2796}/W_{\rm MgII2803}$ and $W_{\rm FeII2586}/W_{\rm FeII2600}$ ratios (shown in the first two panels) asymptote toward the saturation value with increasing $W_{\rm MgII}$. However, neither of these ratios shows any evolution with redshift. In contrast, the $W_{\rm FeII}/W_{\rm MgII}$ ratio clearly evolves, with the median line ratio decreasing with increasing redshift in {\it all} $W_{\rm MgII}$ bins. Fig.~\ref{saturation} also shows that the higher equivalent width bins also exhibit the evolution with redshift, albeit at lower strength than the lower equivalent width (and less saturated) bins. This is consistent with our expectation that saturation effects will tend to weaken the relationship between $\mathcal{R}$ and the intrinsic \fe/\mg\ abundance ratio of the absorbers, and then that these effects are likely only to dampen (rather than induce) a potential trend in $\mathcal{R}$ with redshift.


\noindent{\it (ii) Multiplicity \& Kinematics:}

If the observed absorption lines are instead the result of blends of multiple unresolved (lower equivalent width) lines, then the individual lines might arise in lower column density clouds and are therefore presumably less saturated. The real situation is likely to be more complicated, with lines forming in clouds with varying kinematic properties. 

High-resolution studies have demonstrated that MgII lines are often made up of a number of components, typically between 2 and 10, with the number of components increasing with $W_{\rm MgII}$ \citep{Churchill2003}. 
However, they also find that the MgII and FeII absorption lines show similar Doppler parameters and correlated column densities and the distribution of $N({\rm FeII})/N({\rm MgII})$ column ratios does not depend strongly on velocity spread \citep{Churchill2003}, nor does the kinematic spread show any evidence for significant evolution with redshift \citep{Mshar2007,Churchill2000}.  
Hence, assuming the FeII and MgII absorption lines are equally sensitive to the variations in the clouds as they evolve, 
the evolution of the ratio $W_{\rm FeII}/W_{\rm MgII}$ for individual absorption systems should change monotonically with the abundance ratio $N_{\rm FeII} / N_{\rm MgII}$, although the magnitude of the change may be dampened by saturation effects.  If, for some reason, the velocity structure or Doppler parameter for these two species differs, due to e.g., differences in ionization structure or incomplete mixing of metal species, this monotonic relation can break down.  
For example, incomplete metal mixing could arise in a delayed Fe-enrichment scenario: if the kinematic spread of Fe-rich material is narrower than that of $\alpha$-enriched gas, this could give rise to a suppressed $W_{\rm FeII}/W_{\rm MgII}$ ratio without it reflecting a true dearth of Fe atoms.  However, even if this somewhat contrived scenario were realized, it would still point to an evolution in the sources of CGM enrichment.
While it may be possible to create a trend in $\mathcal{R}$ with redshift as a result of an evolving $b_D$, there is no obvious reason why the Doppler parameter would evolve with redshift for just one species. Better spectroscopic data (higher spectral resolution, and on more species and ionization states) on much larger samples than currently exist can help explore this issue further.

\noindent{\it (iii) Ionization Correction:}
The first two ionization potentials for Mg and Fe are very similar: for MgI$\rightarrow$MgII and FeI$\rightarrow$FeII, they are 7.646 eV and 7.902eV respectively; the second ionization potentials (i.e., MgII$\rightarrow$MgIII and FeII$\rightarrow$FeIII) are 15.0357 eV and 16.1878 eV respectively \citep{Morton2003,NISTASD}. The ratio of the ionic abundances of the two elements, $N_{\rm MgII}/N_{\rm FeII}$, is determined by the shape (at these energies) of the spectrum of the ionizing metagalactic radiation, which does not vary significantly over the redshift range considered here. Using the \citet[]{HM12} model for the UV metagalactic flux as a function of redshift, we find that the ratio of the emissivity at the Mg and Fe 1st ionization potentials varies by less than 4\% for $0.4<z<2.0$; the emissivity ratio at the wavelengths of the 2nd ionization potential varies by less than 2\%. 

The bulk of the absorbers ($>$96\%) in the JHU-SDSS Metal Absorption Line Catalog have $W_{\rm MgII2796}> 0.5$\AA. These absorbers therefore correspond to regions with hydrogen column densities of ${\rm log}(N_{\rm HI})>19.4$ \citep[if the Doppler parameter is 10~km~s$^{-1}$;][]{Menard2009}.  
\citet[]{Giavalisco2011} calculate the variation of the N(MgII)/N(FeII) ratio with ionization parameter for the \citet[]{HM96} metagalactic radiation field. They find that for hydrogen column densities ${\rm log}(N_{\rm HI}/{\rm cm^{-2}})>19$, the N(MgII)/N(FeII) ratio (and hence its inverse) is essentially constant over a wide range of ionization parameter \citep[see also][]{Crighton2015}. 

It is therefore unlikely that differences in ionization could drive the observed trend. 

\noindent{\it (iv) Depletion:}
\mg\ absorption line systems are known to contain dust, exhibiting dust-to-gas ratios comparable to that observed for normal galaxies \citep{Menard2012}. Since both Mg and Fe can be depleted onto dust grains, it is necessary to understand whether differing depletion rates for the two elements as a function of redshift can result in the trend observed in their equivalent width ratios. 

Based on studies of the interstellar medium in our Galaxy, the depletion
rate for Fe is approximately 2$-$7 times larger than that for Mg \citep[e.g.,][]{Savage1996}, i.e., Fe is depleted from the gas phase at a higher rate than is Mg. Using the tabulated model of \citet{Jenkins2009} for the dependence of the Mg and Fe gas phase abundance on the line-of-sight depletion strength $F_*$, we find that the column density ratio ${\rm log}(N_{\rm Mg}/N_{\rm Fe})$ varies as $(0.29\pm0.06)F_*$, i.e., the gas-phase column density ratio increases with increasing depletion and dustier absorbing clouds (which would presumably be more depleted in Mg and Fe) would exhibit higher values of ${\rm log}(N_{\rm Mg}/N_{\rm Fe})$. Hence, the only way to reproduce a trend where the $W_{\rm FeII}/W_{\rm MgII}$ ratio decreases with redshift through purely dust depletion effects is to have higher redshift absorbers be dustier (and more depleted) than lower redshift absorbers, an unlikely scenario.

In contrast, we expect Fe and dust content to increase as the Universe ages and that Fe is likely to be fractionally more depleted onto dust at lower redshift. Hence, any detection of an increase (to lower redshift) in the observed gas-phase Fe/Mg abundance ratio is likely to reflect a stronger intrinsic increase in the abundance ratio. In other words, a measurement of the gas-phase Fe/Mg ratio would underestimate the true Fe/Mg, and the factor by which it is underestimated would be larger at low redshift than at high redshift. This goes in the sense, therefore, of making the observed relation imply a stronger underlying variation. It is therefore unlikely that depletion effects could explain away the observed trend.

\subsection{\it A Simple Physical Model}

In contrast to various possible (and somewhat contrived) explanations discussed above, the expected variation of the rates of SNIa and CCSN should naturally produce the observed variation in $\mathcal{R}$ and we therefore consider this next. 

We model the average universal Fe/Mg abundance ratio as resulting from the supernovae formed at each epoch as a result of the cosmic star formation rate history.\footnote{We ignore the contribution of AGB stars here because (a) supernovae dominate the metal enrichment, and (b) we are considering absorption line systems which likely are clouds at large distances from the galaxy nucleus or disk, where the enrichment is probably due to energetic SN ejecta.} We assume that the rate of core collapse supernovae follows the cosmic star formation rate density, i.e., $r_{\rm CCSN}\propto\psi(z)$ \citep[e.g.,][]{MD2014}. Since CCSN progenitors are massive ($M>8M_\odot$) stars, the rate of CCSN is closely tied to the cosmic star-formation rate density in the Universe, i.e., 
\begin{equation}
R_{\rm CC}(z)=0.0068\psi(z)\,{\rm yr^{-1}\,Mpc^{-3}},
\end{equation}
where $\psi(z)$ is the cosmic star formation rate density \citep[in units of ${\rm M_\odot\,yr^{-1}\,Mpc^{-3}}$;][]{MD2014}. $R_{\rm CC}(z)$ assumes a Salpeter initial mass function with $8<M<40$~\Msun \citep[]{MD2014}.
In contrast, SNIa rates exhibit a ``delay time distribution'' (DTD) relative to the time since star formation \citep[e.g.,][]{Totani2008,Maoz2012a,Maoz2012b,Claeys2014}. Since there is much debate about the exact form of this distribution, here we assume the simple power-law form suggested by \citet{Maoz2012a}: 
\begin{equation}
R_{\rm Ia}(t) = \int_0^t \psi(t-\tau) {\rm DTD(\tau)d\tau}\,{\rm yr^{-1}\,Mpc^{-3}},
\end{equation}
where ${\rm DTD(\tau)=4\times10^{-13}(\tau/1Gyr)}{\rm \, yr^{-1}M_\odot^{-1}}$.
 
We compute the evolving mass density of each element as the sum of the SN yields:
\begin{equation}
M_X=Y_X^{\rm SNIa}\int_0^tR_{\rm SNIa}(t)dt + Y_X^{\rm CCSN}\int_0^tR_{\rm CCSN}(t)dt,
\end{equation}
where $Y_X^{\rm S}$ is the average mass yield for metal $X$ (in \Msun) for a SN of type $S$. The number density of element $X$ is $M_X/A_X$, where $A_X$ is the relative atomic mass of element $X$. The SN yields are very uncertain: based on a review of the literature, \citet{Wiersma2009} conclude that calculated nucleosynthetic yields from SNe vary by factors of $\approx$2. Here, we assume that the typical SNIa produces $8.6\times 10^{-3}$\Msun\ of Mg and 0.74\Msun\ of Fe, and that the typical CCSN produces 0.12~\Msun\ of Mg and 0.09~\Msun\ of Fe \citep{Nomoto1997}. Choosing a single number for the average yields is a simplification, but our aim here is to explore whether the overall variation observed in the $W_{\rm FeII}/W_{\rm MgII}$ ratio can be explained by just the simple evolution of the SN rates.

\begin{figure}
\centerline{\vbox{\hbox{
\includegraphics[width=3.6in]{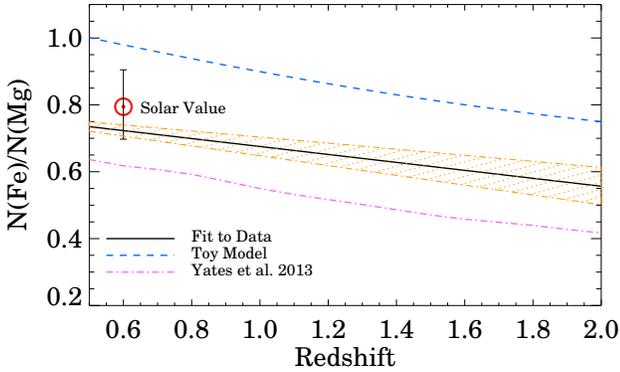}
}}}
\caption{The blue dashed line shows the variation of Fe/Mg abundance ratio with redshift predicted by the simple model described in the text, based on a CCSN rate tied to the Madau \& Dickinson (2014) star-formation rate density and the \citet{Maoz2012a} power-law delay time distribution for the SNIa rate. The dot-dashed purple line shows the variation with redshift of the Fe/Mg ratio computed by \citet{Yates2013} for a model Milky Way - type galaxy. The red circle with the error bar represents the solar abundance ratio. The solid line and the hashed region surrounding it show, respectively, the observed variation in the equivalent width ratio $\mathcal{R}$ and the 99.5\% confidence limits on the slope. We have normalized the fit to match the solar value at $z=0$, under the assumption that the abundances are linearly proportional to the equivalent widths. 
\label{models}}
\end{figure}

The results are shown in Fig.~\ref{models}. The solar abundance ratio of $(N_{\rm Fe}/N_{\rm Mg})_\odot=0.794^{+0.111}_{-0.097}$ is also shown \citep[based on the Mg and Fe photospheric abundances of 7.60$\pm$0.04 and 7.50$\pm$0.04 respectively;][]{Asplund2009}. 
This simple model over-predicts the observed SNIa rates by a factor of 2 and over-predicts the solar photospheric abundance ratio of Fe/Mg at redshift $z=0$ by $\approx40\%$. This is not surprising given the large uncertainties in the yields and our other assumptions (e.g., ignoring the contribution from AGB stars, uncertainties/variations in the IMF, DTD, SFRD, and the large scatter in the data).\footnote{Correcting the $R_{\rm SNIa}$ rate by a factor of 0.6 can result in both a better fit to the observed rates and the solar abundance ratio.} 

Recently, \citet[]{Yates2013} modeled the metal enrichment in a galaxy resulting from AGB stars and SNe, investigating different assumptions for SN yields and SNIa DTDs. They find that the observed local mass-metallicity relationships and the [O/Fe] abundance in the Milky Way can be reproduced by models in which the SNIa DTD is a power-law and the light $\alpha$ elements are expelled in galactic winds. The redshift variation of the Fe/Mg ratio computed by \citet{Yates2013} for a model Milky Way - type galaxy using a power-law DTD is also shown in Fig.~\ref{models}; rather remarkably, its slope matches the evolution in $\mathcal{R}$ observed in the JHU-SDSS dataset. This lends further, though still circumstantial, support that the observed evolution can be explained as the result of the increasing dominance of SNIa yields changing the mean composition of the outflowing gas responsible for the absorption line systems over time.

The comparisons of the models to the slopes measured in the data are only meant to be illustrative and not discriminatory. The scatter in the slope measurement is large and the models are only notional at this stage. Nevertheless it is instructive that the measured variation shows slopes comparable to those predicted by models where all the evolution in the abundance ratios is the result of star formation in galaxies and enriched material ejected into their gaseos halos. The large scatter in the data may reflect the diversity of star formation histories, impact parameters of QSO lines of sight through halos, and inhomogeneities in the halo gas. 

Individual absorbers at any given redshift show significant scatter and the trend (of $\mathcal{R}$ with $z$) reported here is subtle; it was only possible to measure the trend using the very large catalog of absorbers resulting from the SDSS QSO spectra. It is therefore not surprising that previous high-resolution studies \citep[e.g.,][]{Churchill2003,Mshar2007,Narayanan2008,RHidalgo2012} did not see the trend of $\mathcal{R}$ with redshift in the small samples ($<100$ absorbers) studied (however, please see \citet{Srianand1996}'s analysis of the \citet{Steidel1992} dataset, where a possible trend of decreasing $W_{\rm FeII\lambda2382}/W_{\rm MgII}$ is reported). In a VLT/UVES high-resolution study of 100 weak (i.e., $W_{\rm MgII2796}<0.3$\AA) \mg\ absorbers, \citet{Narayanan2008} found that clouds with high $W_{\rm FeII}/W_{\rm MgII}$ disappear from their sample at high redshift ($z>1.2$).
In a complementary study of 81 strong absorbers identified in the same dataset, \citet{RHidalgo2012} found that the $W_{\rm FeII}/W_{\rm MgII}$ ratio for $W_{\rm MgII2796}>1$\AA\ absorbers is typically high and exhibits some evolution with redshift; weaker absorbers (0.3\AA$<W_{\rm MgII2796}<1$\AA) instead showed a wider range and scatter in the $W_{\rm FeII}/W_{\rm MgII}$ ratio with no obvious trend. The authors interpret this as primarily due to kinematics: high-redshift absorbers with $W_{\rm MgII2796}>1$\AA\ tend to be formed in regions with more kinematic spread (and the \fe\ and \mg\ lines are less saturated) than their low-redshift counterparts. They also consider the effects of ionization and $\alpha$-enhancement, but conclude that they cannot discriminate between these effects. 

The SDSS spectra have insufficient resolution to explore this issue adequately \citep{York2000}. However, splitting the sample into two equivalent width bins --- 0.3\AA$\le W_{\rm MgII2796}<1$\AA\ and $W_{\rm MgII2796}\ge1$\AA\ following \citet{RHidalgo2012} --- we find that both subsets show evolution (with slopes and 95\% confidence ranges of 
$-0.0502$, [$-0.0591,-0.0422$] and  
$-0.0434$, [$-0.0516, -0.0344$]
respectively). There is also evidence that the $\mathcal{R}$ values show a much broader distribution for the 0.3\AA$\le W_{\rm MgII2796}<1$\AA\ subset than for the $W_{\rm MgII2796}\ge 1$\AA\ subset. It is not obvious from the large SDSS dataset, therefore, whether splitting the data by equivalent width is justified or if the absorber population should be treated as a continuum, reflecting a range of gas physical conditions in the halos of galaxies. 

Given the various caveats already described in using low-resolution data, it will be necessary to obtain high resolution spectroscopic observations of a sufficiently large sample ($\sim$1000s) of absorbers to properly disentangle the effects of saturation, kinematics and abundance.  A more detailed investigation of the scatter in the equivalent width and ratio distributions compared to the results from numerical simulations will help us further understand the range of physical conditions in galaxy halos.

\section{Conclusion}

We have investigated the equivalent width ratio $W_{\rm MgII}/W_{\rm FeII}$ for a large sample of MgII absorption line systems from the JHU-SDSS Metal Absorption Line Catalog of \citet{ZM13} and discovered that the ratio shows a trend with redshift. This trend does not appear to be the result of data artifacts or selection effects in the catalog. Several reasons may exist for the observed evolution.  However, a simple model shows that the trend is naturally explained by the evolving Fe/Mg abundance ratio that results from the changing relative rates of Type Ia and core-collapse SNe.  Star-formation (and hence supernovae production) is primarily located in galactic disks, whereas the absorption lines under investigation here form in circumgalactic clouds in galaxy halos at large galactocentric distances. Hence, if the equivalent width ratio reflects abundance variations in the CGM clouds, and if the enrichment by supernovae is indeed the primary cause for the CGM metallicity variations, then outflows must be efficient in transporting metals out of star-forming disks to large galactocentric radii. 

Future spectroscopic surveys with the next generation of highly multiplexed multi-object spectrographs (e.g., DESI on the Mayall telescope\footnote{http://desi.lbl.gov/}, PFS on Subaru\footnote{http://pfs.ipmu.jp/factsheet/}, and 4MOST on VISTA\footnote{http://www.aip.de/en/research/research-area-ea/research-groups-and-projects/4most}) will result in larger samples of absorption line systems and the ability to explore the chemical evolution of the Universe in greater detail. Higher signal-to-noise ratio and higher resolution spectroscopy of large samples spanning a large redshift range can confirm whether the trend observed in the Fe/Mg ratio is observed in other $\alpha$-element lines and explore whether the changes are due to the chemical enrichment history of the Universe (as suggested here) or if evolution in the cloud properties (e.g., ionization, kinematics, Doppler parameter, substructure, etc.) is instead responsible for the trend. Detailed comparisons between the data and state-of-the-art cosmological simulations \citep[e.g.,][]{Illustris} will be critical to this effort. 

\section*{Acknowledgements}

This research made use of the JHU-SDSS Metal Absorption Line Catalog \citep{ZM13}, which is based on the SDSS DR7 release \citep{sdssdr7}, and the NIST Atomic Spectra Database \citep{NISTASD}. We thank Dr. Brice M\'enard for useful advice and for putting together such a scientifically useful resource. AD thanks Joan Najita for assistance with the ASURV package and Tom Matheson for useful comments on the manuscript. We are grateful to the referee for a constructive report that resulted in improving this paper. 
Funding for the SDSS and SDSS-II was provided by the Alfred P. Sloan Foundation, the Participating Institutions, the National Science Foundation, the U.S. Department of Energy, the National Aeronautics and Space Administration, the Japanese Monbukagakusho, the Max Planck Society, and the Higher Education Funding Council for England. The SDSS Web Site is http://www.sdss.org/.
%
%
GBZ acknowledges partial support provided by NASA through Hubble Fellowship grant \#HST-HF2-51351 awarded by the Space Telescope Science Institute, which is operated by the Association of Universities for Research in Astronomy, Inc., under contract NAS 5-26555. 
AD's research activities are supported by the National Optical Astronomy Observatory (NOAO). AD thanks the Radcliffe Institute for Advanced Study and the Institute for Theory and Computation at Harvard University for their generous support during the period when this paper was written. NOAO is operated by the Association of Universities for Research in Astronomy (AURA) under cooperative agreement with the National Science Foundation.

\end{document}